\documentclass[10pt,a4paper]{article}
\usepackage[utf8]{inputenc}
\usepackage{amsmath}
\usepackage{amsfonts}
\usepackage{amssymb}

\usepackage{graphicx}
\usepackage{cite}
\usepackage{float}
\usepackage{authblk}

\font\myfont=cmr12 at 14pt

\title{ \myfont On the role of radiative losses in energy scale of large liquid scintillator and water Cerenkov detectors } 

\author{Andrey Formozov \footnote{andrey.formozov@unimi.it}}

\affil{ \fontsize{9}{10.8}\itshape Dipartimento di Fisica, Universita degli Studi e INFN, 20133 Milano, Italy }
\affil{ \fontsize{9}{10.8}\itshape Lomonosov Moscow State University, 119234 Moscow, Russia}
\affil{ \fontsize{9}{10.8}\itshape Joint Institute for Nuclear Research, 141980 Dubna, Russia}

\date{}

\begin{document}
\maketitle

\begin{abstract}
The scope of this article is to understand the role of radiative losses in detector energy scale  at energies of 5-100 MeV and to estimate the non-linear effect they produce in large liquid scintillator and Cherenkov detectors.  
For this purpose the generalized energy scale detector model was constructed and the detector response was simulated by means of toy Monte Carlo simulations. It was found that the non-linearity grows proportionally to a relative fraction of the Cherenkov light into the total light yield achieving the maximum for pure water Cherenkov detectors. The effect generally increases with energy and then saturates at higher energies. In the energy region 5-10 MeV for all types of detectors and for energies 5-100 MeV in scintillation detectors with low fractions of Cherenkov light ( $ < $ 10 \% )  the non-linearity is estimated to be a sub-percent effect. Meanwhile it could reach the order of 5\% for 50 - 100 MeV electrons in pure water Cherenkov detectors. It was hypothesised that this effect can contribute to the difference in observed reactor neutrino spectra. The present analysis showed that the effect was quite small to generate significant differences in observed reactor neutrino spectra. Though it still may be relevant for precision measurements and phenomenological analysis in aforementioned energy ranges. 
\end{abstract}

\section*{Introduction}

The detection and analysis of rare neutrino events with energies between 5  and 100 MeV require a good understanding of the detector energy scale at high energies, since direct calibrations in this energy range are rarely available. This energy range is relevant for the study of reactor neutrino ( upper part of the spectrum), solar neutrinos and neutrino detection from supernova and core-collapsing stars \cite{cs_vissani}.

A particularly interesting problem is a shape distortion in the 4-6 MeV region of reactor neutrino spectrum observed by several experiments at the same time. In the article  \cite{mention2017reactor} dedicated to this problem, it was shown in  that the reactor neutrino spectra of Bugey 3 \cite{achkar1996comparison}, Daya Bay \cite{An:2015rpe}, Double Chooz \cite{abe2014improved} and RENO \cite{Seo:2016uom} experiments are not compatible and that it may be related to the energy scale distortion.

Two of the most common reactions in large-scale liquid scintillator detectors are neutrino inverse beta decay with a positron in the final state \cite{cs_vissani} and elastic neutrino-electron scattering \cite{Phase1}. The second reaction is also very common in water Cherenkov detectors \cite{Fukuda:2002pe}.

The final state electron (positron) \footnote{  Here we consider only electrons, since the energy scales of electrons and positrons are essentially the same apart from the fact that the energy scale of positrons is shifted by two annihilation gammas} provides information about the neutrino energy and momentum. Therefore, with precise measurement of the electron spectrum the reconstruction of the initial neutrino spectrum is possible. For this purpose the detector non-linearity of the energy scale must be well understood, usually by means of calibrations, Monte Carlo simulations and analytic descriptions \cite{Agostini:2017aaa}, \cite{Phase1},\cite{an2014spectral}.

While the nonlinear effects of the energy scale at low energies are usually very well described and estimated, the effects of the non-linearity at higher energies have not yet been explicitly discussed.

Radiative energy losses may play a significant role in energy scale nonlinearity at higher energies, since in this case the event comprises several secondary radiated gammas and parent electrons and the response of the detector to electrons and gammas could be different.

The goal of this article is to estimate the role of radiative losses in detector energy scale non-linearity at high energies and to estimate the non-linear effect they produce in liquid scintillator and water Cherenkov detectors. For this purpose the generalized detector energy scale model was built and Monte Carlo simulations were conducted. To consider the problem in general for a wide range of different detectors the difference in the energy response between them is parameterized by varying contribution of the Cherenkov light to the total energy estimator ( also know as visible energy or  total light yield ).

\section*{Formalism of the idea}  

\begin{figure}
 \centering
   \includegraphics[width=0.8\linewidth]{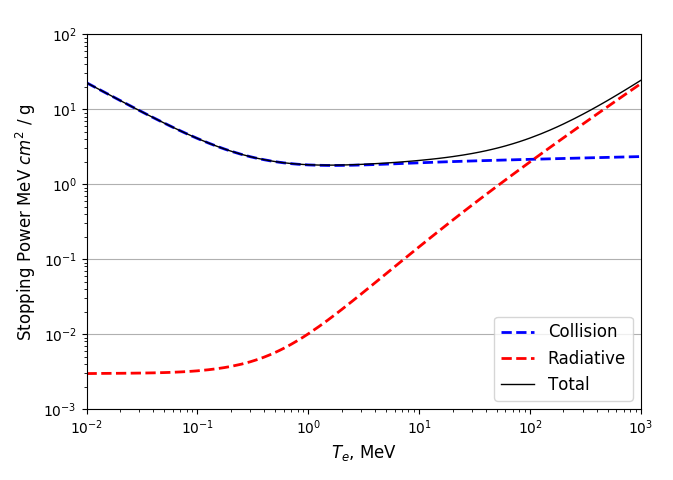}
 \caption{Ionization (blue dotted line) and radiative (red dotted line) stopping powers in benzene that determine energy losses of electrons. At low energies ($< 1 \text{MeV}$) ionization losses are maximal. In the middle and high range they are almost constant. At higher energies radiative losses become relevant. Adopted from \cite{nist}.}  
 \label{fig:stoppingpower}
\end{figure}  

Incident neutrinos could be reconstructed from the products of reactions of neutrinos with matter, where in most cases an electron (positron) is a final-state particle. The incident electron loses its energy in the liquid scintillator through collisions with other electrons, which causes ionization of the media ( fig. \ref{fig:stoppingpower} ). The energy of electron is not observed directly but rather estimated by means of some physical process: collection of the ionization charges, amount of scintillation light and/or Cherenkov emission, and other processes. This measurable quantity will be referred to as the estimator.

For liquid scintillator and Cherenkov detectors, one can choose the total light yield ( number of registered photons ) as an estimator $E_{vis}$ ( visible energy ). The relation between the energy of an incident particle $E_{true}$ and the energy estimator is determined by the detector energy scale that, in principle, could be parameterized analytically. In case of an ideal detector energy scale the estimator is strictly proportional to the energy. 

In the real case, several effects make energy estimation more complicated and electrons with different energies have different coefficients of proportionality with the estimator: the model became nonlinear.

One may construct an analytical model $E_{vis}(E_{true})$ of the energy scale of the liquid scintillator or Cherenkov detector incorporating the effects of the quenching and Cherenkov  emission.  Such a model is of high value for the analysis when direct Monte Carlo simulations of the detector are not available (e.g in a phenomenological analysis). 

Here it is demonstrated that Monte Carlo simulations are to some extent unavoidable if one would like to apply the model for the events at higher electron energies (5-100 MeV). A very distinct process of energy losses - radiative losses or bremsstrahlung - becomes relevant, since the collisions of electrons with other electrons and nuclei change the direction of the electron, resulting in emission of "hard" photons (fig. \ref{fig:stoppingpower}).

Thus, the total energy deposited by electron in media can be divided into two parts:

\begin{equation}
E_{true}^{e} = E_{ion} + E_{rad} = E_{ion} + \sum_i E_{i, rad}^{\gamma}
\end{equation}

Electrons and gammas release their energy in significantly different ways. In liquid scintillator and water detectors gammas produce {\em less} or equal amount of light than electrons:

\begin{equation}
E_{vis}( E^{gamma} ) < E_{vis}( E^{e} )
\end{equation}

 As a consequence, they have different energy scales and the contributions of $E_{ion}$ and $E_{rad}$ into the energy estimator $E_{vis}$ are significantly different. Since the contribution from radiation losses increases with energy, the effect becomes more and more relevant as energy increases.  In this article,  the effects of radiation losses on the non-linearity of the energy scaling at higher energies is explored.

For these energies, to find a value of the estimator one needs to treat electrons and radiative gamma separately and the $E_{vis}$ is given by expression:

\begin{equation}
E_{vis} =  E_{vis} ( E_{ion}^{e} ) + \sum_i E_{vis}( E_{i , rad}^{\gamma} )
\end{equation}

which must be less than $E_{vis}(E_{true}^{e})$.


\section*{Energy model}

An analytical energy model could be constructed incorporating scintillation quenching and Cherenkov radiation effects.

Applying the model the total amount of light (collected charge) for an electron of a given energy can be obtained:

\begin{equation}
E_{vis}(E) =LY \: f_{quench}(E,kB) \: E + f_{cher}(E,n)
\end{equation}

where $LY$ - is the scintillator light yield, $f_{quench}(E , kB)$ is the quenching factor and  $f_{cher}(E , n)$ is the Cherenkov contribution factor, which will be discussed further in detail. Note that these factors are detector-dependent. To include this dependency in the model we scale the function $f_{cher}( E, n )$ with a factor $\alpha_{cher}$ to increase the relative contribution of the Cherenkov light into the total light yield ( energy estimator $E_{vis}$):

\begin{equation}
E_{vis}(E) =LY \: f_{quench} (E, kB) \: E + \alpha_{cher} f_{cher}( E, n )
\end{equation}

Varying $\alpha_{cher}$ we can study the whole range of liquid scintillator detectors, which due to different scintillator compositions have different Cherenkov contributions. Here the role of the quenching effect is not considered explicitly and $LY \: f ( E, kB )$ is kept fixed. 

As a measure of the importance of the Cherenkov effect one can define a value of the relative contribution of Cherenkov light at 1 MeV:

\begin{equation}
\eta_{cher} = \alpha_{cher} \: f(1 \: \text{MeV}, n) \: / E_{vis}(1 \: \text{MeV})
\end{equation}

In the extreme case of a pure Cherenkov detector ($\eta_{cher} = 1$, $\alpha_{cher} \rightarrow \infty$ ) the model simplifies to:

\begin{equation}
E_{vis} = f_{cher}(E,n) 
\end{equation}

More detailed descriptions of the energy model may be found in \cite{Phase1}.

\section*{Interaction of the gamma rays with the media}

The way in which gamma rays interact with the media differs significantly from charged particles. It is determined by the photoelectric effect, Compton scattering, pair production \footnote{ Here we do not consider the nuclear photoelectric effect} \cite{Knoll}. 

The photoelectric effect is the process of interaction of a photon with one of the internal electrons of the atom when all the energy of the photon is transmitted to the electron. In liquid organic scintillators and water it is relevant for energies below 100 keV.  With increasing gamma energythe Compton scattering process becomes more relevant for energy losses andn the energy region 0.1-1 MeV it is the main process. In large liquid scintillation detectors such gamma quanta usually experience several Compton scattering processes until they reach the energy range of the photoelectric dominance.

The threshold of electron-positron pair production is at 1.02 MeV and its cross section grows logarithmically with energy.  The electron and positron deposit their energy in the liquid scintillator. 
After positron annihilation, 0.511 MeV gammas deposit their energy though the Compton and photoelectric effect described above. 

Bringing all this together, the energy deposited by gammas could be written as a sum:

\begin{equation}
E^{\gamma} = \sum_j E_{j, comp/phot/pair}^{e}
\end{equation}

The energy estimator for a gamma may be represented as
 
\begin{equation}
E_{vis}(E^{\gamma}) = \sum_j E_{vis}(E_{j, comp/phot/pair}^{e})
\end{equation}

\begin{figure}
 \centering
   \includegraphics[width=0.7\linewidth]{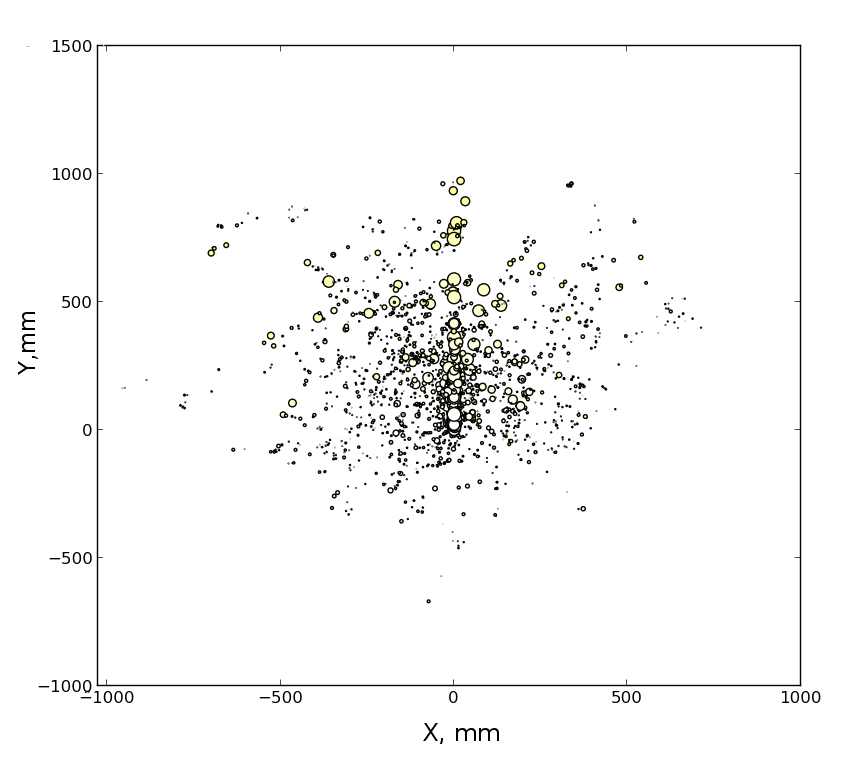}
 \caption{Monte Carlo simulations. The vertices of interaction of 1 MeV 1000 gamma particles in a liquid scintillator. The initial momentum is along the y-axis. The starting point is at the origin. The size of circles is proportional to deposited energy. } 
 \label{fig:gamma}
\end{figure}

The total length of the track of the gamma quanta with energy 1 MeV in a liquid scintillator could be up to 1.5 m long but Compton scattering, changing the direction of the gamma particle, significantly reduces the effective distance of propagation. Fig. \ref{fig:gamma} illustrates the vertices of interaction of 1 MeV gamma particles in a liquid scintillator. The majority of events could be located in a sphere with radius 0.5 m. The spatial topology of the gamma event significantly differs from the electron point-like event. Note that gamma produced at the border of the detector can actually escape from it.

\section*{Quenching} 

Ionization quenching is the dependency of the light yield  from the differential energy losses \cite{Birks}:

\begin{equation}
dL \propto dE/ (1+kB \: dE/dx)
\end{equation}

, where $dL$ is the number of photons per unit particles energy,  $dE/dx$ is the ionization stopping power of the particle (fig. \ref{fig:stoppingpower}) and $kB$ is a quenching constant.  

Integrating this relation and normalizing it to the particle energy one can obtain the quenching factor:

\begin{equation}
f(E) = 1/E \int_{0}^{E} dE'/ (1+kB \: dE'/dx)
\end{equation}

To build a realistic energy model of liquid scintillator detectors we take $kB = 0.011 \text{cm}  \text{MeV}^{-1}$ and keep it fixed. The plot of $f_{quench}(E,kB)$ for this quenching constant may be found in fig. 8 \cite{Phase1}. 

\section*{Cherenkov effect} 

A charged particle starts to emit Cherenkov light when it propagates with a speed higher than the speed of light in the media \cite{cerenkov1934}. The condition for Cherenkov emission of a photon with a given wavelength $\lambda$ is related to the value of the refractive index $n(\lambda)$ of the media and velocity of the charge particle $\beta=v/c$.

The condition for radiation may be formulated as:

\begin{equation}
\label{CherenkovCondition}
n(\lambda) > 1 / \beta   
\end{equation}

Since $n(\lambda)$ depends on the wavelength of the emission $\lambda$, the threshold for different wavelengths will generally be different.

The spectrum of Cherenkov light could be found by the formula for an average number of photons emitted in the energy interval $\epsilon$ while particle moves the distance $dx$  \cite{tamm1937coherent}, \cite{collaboration2016physics}.

\begin{equation}
\label{form:cherenkovspectrum}
\frac{dN}{dx d\epsilon} = \frac{\alpha z^2}{ \hbar c} (1 - \frac{1}{n^2 \beta^2}) \approx 370 z^2 \frac{photons}{eV cm} (1 - \frac{1}{n^2 \beta^2})
\end{equation}

To find the final emitted spectrum and number of photons one needs to integrate this relation over particle track and photons energy.

\begin{figure}
 \centering
   \includegraphics[width=\linewidth]{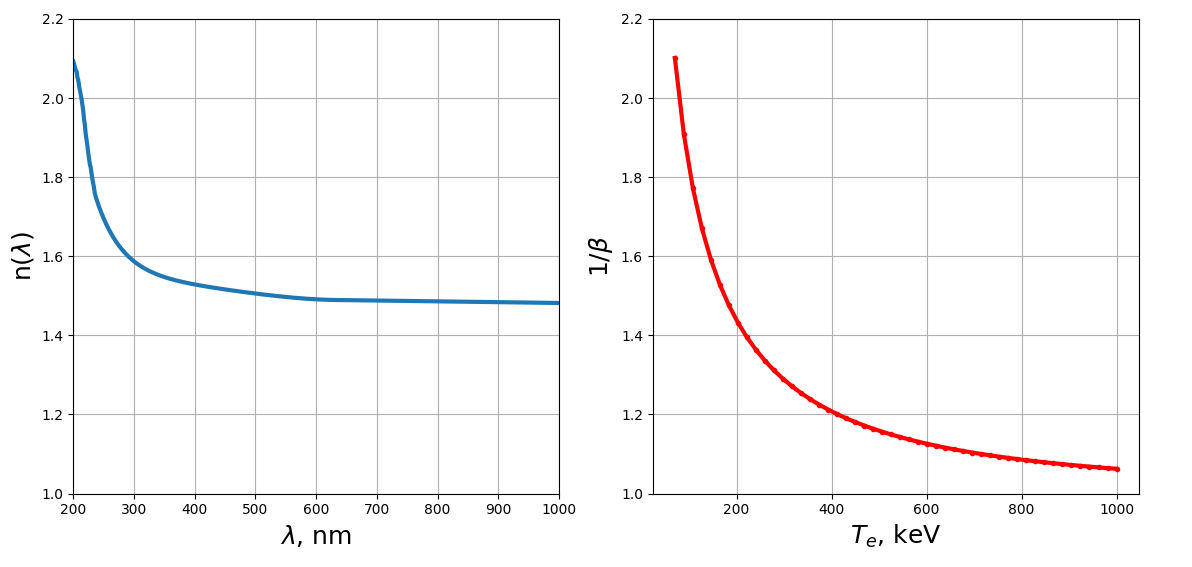}   
 \caption{On cherenkov spectrum.  Left  panel, blue line: the dependency of the refractive index on the wavelength of the photon $\lambda$. Right panel, red line: the dependency of the factor $1/\beta$ from the kinetic energy of electron $T$. Look in the text for explanation.} 
 \label{fig:cher}
\end{figure}

Fig. \ref{fig:cher} shows how the Cherenkov emission spectrum evolves with energy. Two subplots illustrate how the condition \ref{CherenkovCondition} works for different kinetic energy of the particle $T$ and different wave lengths $\lambda$ . On the right panel the red line represents the dependency of the factor $1/\beta$ from the kinetic energy $T$. This plot correspond to the right part of the condition \ref{CherenkovCondition}. The left subplot represents the dependency of the refractive index on the wavelength of the photon $\lambda$ and corresponds to the left part of the condition \ref{CherenkovCondition}.

To obtain the spectrum of Cherenkov emission for a given kinetic energy $T$, one needs to draw a horizontal line that crosses both subplots though the corresponding point $1/\beta(T)$. All points of the left subplot that are above the horizontal line indicate wavelengths of radiation. They satisfy condition \ref{CherenkovCondition}.  The spectrum of the Cherenkov light could be calculated if the measurement for refractive index $n(\lambda)$ as a function of wavelength $\lambda$ is provided. As can be seen the emission actually starts from an energy of about 80 keV for a wavelength about 200 nm (point a) . As the energy increases, the condition \ref{CherenkovCondition} becomes valid for longer wavelengths until the spectrum entirely covers the range 200-1000 nm at approx. 180 keV ( $1/\beta \approx 1.5$), after which the spectrum remains unchanged at higher energies. 

However, based on these calculations it is not possible to state the contribution of the Cherenkov light to the total registered light. This value depends on the further evolution of the Cherenokov photons with a given wavelength $\lambda$ that is determined by the liquid scintillator composition, transparency and quantum efficiency of the light detectors. Moreover, the UV part of the Cherenkov spectrum has not been considered at all for simplicity of the calculations. This fact doesn't have a large influence on the the result of the analysis since the main characteristic parameter is $\eta$.

The Cherenkov factor $f_{cher}(E, n)$ could be found by integrating $\frac{dN}{dx d\epsilon}$ over electron track and radiated photon energy $d\epsilon$.

There are two main factors that determine the nonlinearity of the Cherenkov factor.  

Non-linearity of the Cherenkov factor could be expressed as a derivative $d f/ dE$.

From the point a) until the point b) the spectral range of Cerenkov spectrum is growing, increasing the contribution of the Cherenkov ( $df/dE$ incleases). 

However, there is another even more important mechanism generating non-linearity : the dependency of the $df/dE$ on the ionization stopping power $dE/dx$ (fig. \ref{fig:stoppingpower}). Actually,  

\begin{equation}
df/dE \propto dN_{cher}/dE = \frac{dN_{cher}}{dx}  \frac{dx}{dE}
\end{equation}

therefore $df/dE$ is not constant until $dE/dx$ changes. 
Finally, Cherenkov response $df/dN$ becomes fully linear when the stopping power $dE/dx$ reaches a plateau at about $1.5 MeV$.

\section*{Detector response simulations}

For a good illustration a hypothetical situation is considered when no radiation gammas are registered by the detector. In this case the visible energy obtained will consist only of the ionization part. It is equal to the energy of electron minus the sum of the energy of all radiative gamma emitted:

\begin{equation}
E_{vis}^{w/o \: rad} = E_{ion}^{e} = E_{true}^{e} - \sum E_{i, rad}^{\gamma}.
\end{equation}

\begin{figure}
 \centering
   \includegraphics[width=\linewidth]{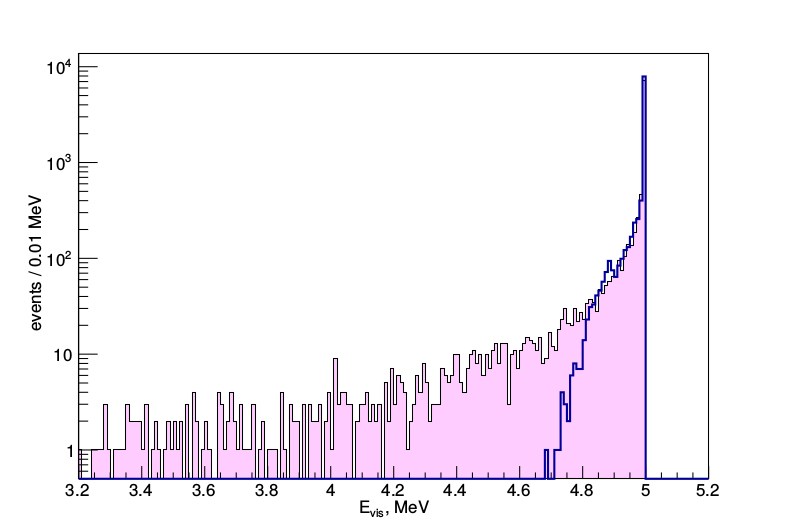}
 \caption{The distribution of 5 MeV 10000 events consisting of ionization energy deposition without radiation gamma contribution (violet histogram) and with the general detector energy model applied (solid blue line).}
 \label{fig:dist5MeV}
\end{figure}

\begin{figure}
 \centering
   \includegraphics[width=\linewidth]{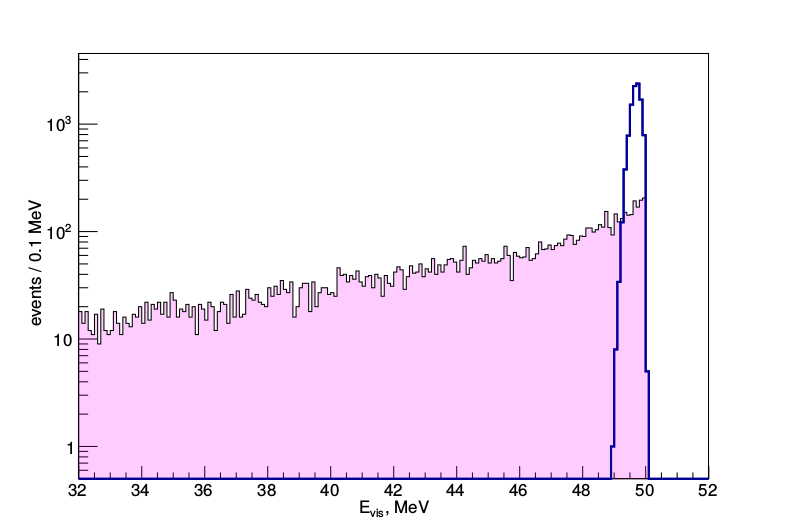}
 \caption{The distribution of 50 MeV 10000 events consisting of ionization energy deposition without radiation gamma contribution (violet histogram) and with the general detector energy model applied (solid blue line).}
 \label{fig:dist50MeV}
\end{figure}

The distribution of the events consisting of ionization energy deposition without radiation gamma contribution are represented by the violet histogram in fig. \ref{fig:dist5MeV} and fig.  \ref{fig:dist50MeV} for 5 MeV and 50 MeV, respectively. The distribution of events has a long tail. The effect is stronger for higher electron energies, as is expected since radiation gammas became more energetic and frequent. In a detector the exact structure of this distribution is hidden by energy resolution and what remains relevant is the shift of the mean of the distribution. The difference between the mean of the distribution and the true electron energy increases, forming the nonlinear response at higher energies.

 In the real case the energy of the radiated gamma is also deposited in the detector and it contributes to the total energy estimator (visible energy), although a gamma makes a smaller contribution to the total energy estimator than an electron of the same energy.  The biggest suppression could be observed in the Cherenkov detector, where only a high energetic gamma that has secondary particles with an energy more than the threshold of the Cherenkov light production could be observed \footnote{Moreover, Cherenkov photons produced by Compton events do not contribute to the cone of the primary particle}.

In this case, the energy estimator for a given electron energy may be calculated as a sum of the estimator for each secondary electron of each radiated gamma and ionization part of the electron:

\begin{equation}
E_{vis}  = E_{vis}( E_{ion} ) + \sum_{i,j} E_{vis}(E_{ij, comp/phot/pair}^{e})
\end{equation}

where index $i$ is a summation over the radiative gamma and $j$ is a summation over all secondary particles.

The normalization of the energy model is chosen such that $E_{vis}( E_{true} ) = E_{true}$ for the events with no radiative losses, when the entire electron energy is deposited by ionization.

Since the energy of the radiative gamma becomes visible, the effect of the smearing and the shift of the energy distribution becomes less dramatic, though still significant. 

For the illustration of more realistic case the general detector energy model is considered with a 10 \% relative contribution of the Cherenkov light at 1 MeV.

 The distribution of $E_{vis}$ events is represented in fig. \ref{fig:dist5MeV} and fig. \ref{fig:dist50MeV} by solid lines. For the 50 MeV electrons a shift $dE = 0.4 \: \text{MeV}$ of the mean of the distribution is visible, which corresponds to 0.8 \% of the non-linearity of the energy scale. 
 
 \section*{Results and conclusions} 

\begin{figure}[H]
 \centering
   \includegraphics[width=\linewidth]{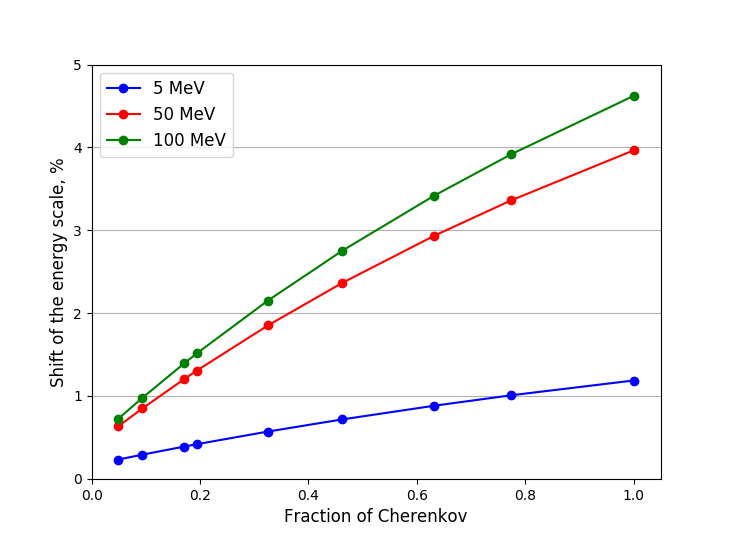}
 \caption{The dependency of the relative energy scale shift $dE = ( <E_{vis}> - E_{true} ) / E_{true}$ in \% as a function of the fraction of Cherenkov $\eta$ for electrons of energy 5 MeV, 50 MeV and 100 MeV.}
 \label{fig:RelContribution}
\end{figure}

Now let's analyse the problem more generally. In  fig. \ref{fig:RelContribution} the dependency of the $dE = ( <E_{vis}> - E_{true} ) / E_{true}$ as a function of the relative Cherenkov contribution $\eta$ is shown for electrons of energy 5 MeV, 50 MeV and 100 MeV.

The effect grows almost linearly with fraction of Cherenkov light $\eta$ into the total light yield. As expected, the greater the energy of electron, the stronger the non-linearity. Surprisingly, at higher energies we can see the effect of saturation on the shift (the curves for 50 MeV and 100 MeV are very close to each other): the energy scale becomes linear again. Careful analysis of the topology of events shows that this is due to an increase in electron-positron pair production in gamma energy losses. A pair deposits its energy via ionization, reducing the role of radiative energy losses in non-linearity formation.  

Typical Cherenkov fraction in liquid scintillator detectors is less than 10 \%. From the fig.
\ref{fig:RelContribution} it is clear that for this fraction at the energy 5 MeV the discussed effect is sub-percent. Therefore, it can not significantly contribute to the difference in observed reactor neutrino spectra, which is order of 1 \% \cite{mention2017reactor}. Nevertheless, the effect may be still relevant for precision measurements and phenomenological analysis in aforementioned energy ranges.




\bibliography{biblio}
\bibliographystyle{unsrt}

\end{document}